\documentclass{amsart}
\usepackage{amsfonts}
\usepackage{amssymb,amsthm,amsmath}
\usepackage{graphicx}
\usepackage{graphics}
\usepackage{subfigure}
\usepackage{url}
\usepackage[english]{babel}
\usepackage{multirow}
\usepackage{longtable}
\usepackage{caption}
\usepackage{amsfonts} 

\usepackage{amsthm}

\usepackage{amssymb}
\usepackage{color}
\usepackage{float}
\restylefloat{table}
\usepackage{titletoc}
\usepackage{rotating}

\usepackage{amsthm}  

\usepackage[toc,page,titletoc,title]{appendix}

\usepackage[english]{babel}

\newcommand{\im}{\text{i}}
\newcommand{\R}{\mathbb{R}}

\newtheorem{theorem}{Theorem}[section]

\newtheorem{example}[theorem]{Example}

\begin{document}

\title{Pricing Energy Contracts under Regime Switching Time-Changed models}%
\author{Konrad Gajewski and Sebastian Ferrando and Pablo  Olivares}%



\begin{abstract}
The shortcomings of the popular Black-Scholes-Merton (BSM) model have led to   models which could more accurately model the  behavior of the underlying assets in energy markets, particularly in electricity and future oil prices. In this paper we consider  a class of regime switching time-changed Levy processes, which builds upon the BSM model by incorporating jumps through a random clock, as well as randomly varying parameters according to a two-state continuous-time Markov chain. We implement pricing methods based on expansions of the characteristic function as in \cite{Fourier}. Finally, we estimate the parameters of the model by incorporating historic energy data and option quotes using a variety of methods.
\end{abstract}
\maketitle
\section{Introduction}
A main goal of the paper is to develop a methodology to price  European option's contracts on electricity and future oil prices.  The approach is based on Fourier expansions and implements models that capture specific stylized features of the underlying assets such as stochastic volatility and random jumps. In particular, we consider a switching time-changed Levy process as an alternative to the BSM model and implement a pricing algorithm based on the expansion of its characteristic function as considered in \cite{Fourier}.\\

 We compare the prices we obtain with those obtained via a computationally costly, but accurate, Monte Carlo method and study sensitivities with respect to relevant parameters, e.g. maturity, strike price and initial price. Moreover, we contrast prices under the regime switching model to those given by the Black-Scholes equation and show that the prices agree when the switching model is reduced to the Black-Scholes model.\\

 We rely on the Esscher transformation, see \cite{EsscherTransform}, to obtain
an equivalent martingale measure(EMM) and  in order to work on a risk-neutral setting.  To calibrate parameters, we use market option prices and minimize the mean squared error. On the other hand, and in order to estimate parameters, we implement  the method of moments, minimum distance estimation and maximum likelihood estimation techniques. For simplicity an Expectation Maximization algorithm (EM) is not considered.\\

 Although most of these elements have been previously implemented, to the best of our knowledge, the combination consisting of the selected model class, the pricing methodology, the risk-neutral framework and the estimation/calibration approach have not been studied before in energy markets or elsewhere. It is worth noticing that non-switching models with Levy noises have been introduced in \cite{benth1}. On the other hand, results for switching Levy models and their characteristic functions can be found in \cite{Switching}.\\

 Many financial time series, including commodity futures seem to exhibit dramatic breaks in their behaviour, for example in the events of political changes or financial crises. Different intervals sharing similar characteristic can be grouped together under a single regime. Models that can capture such behaviour are  regime switching Levy.  Under such a model, the process switches randomly between different Levy processes according to an unobservable Markov chain. The regime switching time-changed  Levy process is a pure jump process which captures two key features of the market: the existence of regimes and price jumps.\\

The organization of the paper is the following:
Section \ref{modelsAndCharacteristicFunctions}  introduces the  regime switching time-changed Levy model, we then derive its characteristic function under Gamma and Inverse Gaussian subordinators. In section \ref{pricingMethods} we discuss   Monte Carlo and Fourier Cosine pricing methods. For Monte Carlo, we develop an algorithm to simulate trajectories of the process, as well as for pricing European call options by simulating many regime switching processes simultaneously. In section \ref{estimationAndOutput} we use  calibration and various  methods to estimate values of model parameters from option quotes and historical prices of oil and electricity commodities.

\section{Model, contract  and characteristic function} \label{modelsAndCharacteristicFunctions}
 Let  $(\Omega ,\mathcal{A}, (\mathcal{F}_{t})_{t \geq 0}, P)$ be a filtered probability space verifying the usual conditions. For a stochastic process $(X_t)_{t \geq 0}$ defined on the space filtered space above the functions $\varphi_{X_{t}}(u)$ and  $\Psi_X(u)=\frac{1}{t} \log \varphi_{X_{t}}(u)$   define its characteristic function and characteristic exponent  respectively. When the process has stationary and independent increments the later does not depend on $t$.
  $A^T$ is the transpose of matrix $A$ unless it is specified differently.\\
 Let the process $\{S_t\}_{t\geq 0}$ represent the price of the underlying asset at any time $t>0$, $X_t=\log S_t$ represents the logarithm of the prices.\\
We define a continuous-time Markov chain $\{s_t\}_{0\leq t\leq T}$ driving the changes between regimes with the state space  $E=\{1,2\}$. The switching times are described by a sequence of  independent random variables $(\tau_j)_{j \in \mathbb{N}}$ in a way that:
    \begin{equation*}
    \lim_{t\rightarrow\tau^-_k}s_t\neq s_{\tau_k}\:\:\:\: \text{for}\:\:\:\:k=1,2.
\end{equation*}
 The infinitesimal generator matrix of the continuous-time Markov chain is given by
 \begin{equation}
  Q=\begin{bmatrix}\label{Q}
-\lambda_{12} & \lambda_{12}\\ \lambda_{21} & -\lambda_{21}
\end{bmatrix}.
\end{equation}
Hence:
\begin{eqnarray*}
  P \{s_{t+h}=2/s_t=1 \} &=&  \lambda_{12}h+o(h)\\
   P \{s_{t+h}=1/s_t=2 \} &=&  \lambda_{21}h+o(h)
\end{eqnarray*}
Next, consider a collection of independent subordinators $L^j=\{L^j_t\}_{t \geq 0}$ for $j\in E$, where each subordinator $L^j$ is also independent of each process $X^i$, for $i, j\in E$.  Each subordinator is characterized by two parameters $\alpha_j, \beta_j >0$ which change between states on the (non-observable) Markov Chain. Each process $L^j$ is a pure jump process and each process $X^j$ has almost surely continuous paths \\
We define the collection of time-changed Levy processes    $Y^j=\{Y_t^j\}_{t \geq 0}$ where
 \begin{equation*}\label{time_changed_brownian_motion}
    Y_t^j=\mu_j L_t^j+\sigma_j B_{L_t^j}.
\end{equation*}
with $\mu_j\in\R$ and $\sigma_j>0$.\\
There exists a natural  economic interpretation of a time-changed process. Energy markets alternate at random times between calmed and frenzy periods at random times.\\
 We now define the regime switching time-changed Levy process $Z=\{Z_t\}_{t \geq 0}$ as:
\begin{equation} \label{switch}
    Z_t\equiv Y_t^{s_t}\:\:\:\:\text{where}\:\:\:\: Y^{s_t}_t=\mu_{s_t}L^{s_t}_t+\sigma_{s_t}B_{L^{s_t}_t}
\end{equation}
or, in differential terms:
\begin{equation*}
  dZ_t\equiv dY_t^{s_t}\:\:\:\:\text{where}\:\:\:\: dY^{s_t}_t=\mu_{s_t}dL^{s_t}_t+\sigma_{s_t}dB_{L^{s_t}_t}
\end{equation*}
The regime switching time-changed Levy process $Z$ is assumed to be the log-price process of the underlying asset and the stochastic process of the asset price itself $\{S_t\}_{t \geq 0}$  is defined as:
\begin{equation*}\label{stockpriceprocess}
    S_t=S_0\exp(Z_t).
\end{equation*}
For simplicity we assume that the process will always start out at state 1 with probability 1.\\

Following  \cite{Switching}, for  the process $Z$ defined in equation (\ref{switch}) along with a  Markov chain with the infinitesimal generator matrix $Q$ defined in equation (\ref{Q}), its characteristic function is given by:
\begin{equation}\label{eq:charfunc}
  \varphi_Z(u)=\exp(i u y_0)~ \mathbb{E}_{\mathcal{Q}}\{[1, 1] \exp(t \Phi(u))[1,0]^T \}
\end{equation}
where $y_0= \log S_0$ and $\Phi(u)$ is the matrix:
 \begin{equation*}
   \Phi(u)=\left(
                     \begin{array}{cc}
                      -\lambda_{12}+ \Psi_{L_t^{(1)}}(\mu^1 u +\frac{1}{2}i \sigma^2_1 u^2)   &  \lambda_{12}\\
                       \lambda_{21} & -\lambda_{21}+ \Psi_{L_t^{(2)}}(\mu_2 u +\frac{1}{2}i \sigma^2_2 u^2)  \\
                     \end{array}
                   \right).
\end{equation*}
 Notice that conditionally on $s_t=j$ the characteristic function of $Z^j$, i.e. the characteristic function of $Z$ conditionally on $s_t=j$,  is:
\begin{equation*}
  \varphi_{Z_t}(u)=\varphi_{L^j_t}(\mu_j u +\frac{1}{2}i \sigma^2_j u^2).
\end{equation*}
 To  compute the exponential matrix $\Phi(u)$ we use a \textit{scaling and squaring algorithm}, see \cite{expm2}, based on the following approximation:
\begin{equation*}
    e^{\Phi(u)}=(e^{2^{-s}\Phi(u)})^{2^s}\approx r_m(2^{-s}\Phi(u))^{2^s},
\end{equation*}
where $r_m(x)$ is the $[m/m]$ Pade approximant of $e^x$ and the nonnegative integers $m$ and $s$ are chosen in such a way as to achieve minimum error at minimal cost. A table of errors as a function of $s $ and $m$ is given in \cite{expm1}. The $[k/m]$ Pade approximant for the exponential function is:
\begin{equation*}
    r_{km(x)}=p_{km}(x)/q_{km}(x)
\end{equation*} where:
\begin{equation*}
    p_{km}(x)=\sum_{j=0}^k \frac{(k+m-j)!k!}{(k+m)!(k-j)!}\frac{x^j}{j!},\:\:\: q_{km}(x)=\sum_{j=0}^m\frac{(k+m-j)!m!}{(k+m)!(m-j)!}\frac{(-x)^j}{j!}.
\end{equation*}
 The discounted price process is denoted $\Tilde{S}=(\Tilde{S}_t)_{t \geq 0}$ where $\Tilde{S}_t:=exp(-rt)S_t$. We have that under an EMM  $\mathcal{Q}$, the discounted price process $\Tilde{S}$ is a martingale under $\mathcal{Q}$ if and only if the following equation is satisfied:
\begin{equation}\label{martingale_condition}
    \Psi_{Z}(-\im)=r.
\end{equation}
 See \cite{Martingale} for details.
\begin{example}\textit{Case of Inverse Gaussian and Gamma subordinators.}\\
Inverse Gamma and Gamma subordinators have been studied in \cite{ig} and \cite{gamma}.\\
 When the subordinator $L^j$ is an Inverse Gaussian process with shape parameter $\alpha_j$ and rate parameter $\beta_j$, we have:
\begin{equation}\label{eq:tc_ig}
    \Psi_{Z^j}(u)=\alpha_j (\sqrt{2 (\mu_j u +\frac{1}{2}i \sigma^2_j u^2)+\beta_j^2}-\beta_j),\alpha_j>0,\beta_j>0, j=1,2.
\end{equation}
In Figure \ref{fig:subim2} the real and imaginary parts of the characteristic function and the characteristic function of $Z=\{Z_t\}_{t\geq 0}$ with an Inverse Gaussian subordinator are shown. Parameters are obtained from estimating procedures for future oil prices as explained in  Section \ref{estimationAndOutput}. A similar result is obtained under a Gamma subordinator.
\begin{figure}[htb!]
\begin{center}
\subfigure[]{
\resizebox*{7cm}{!}{\includegraphics{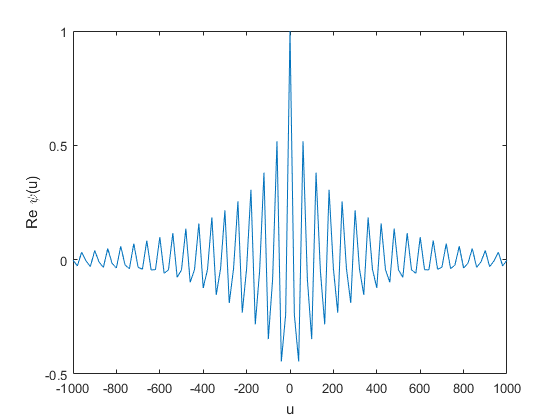}}} 
\subfigure[]{
\resizebox*{7cm}{!}{\includegraphics{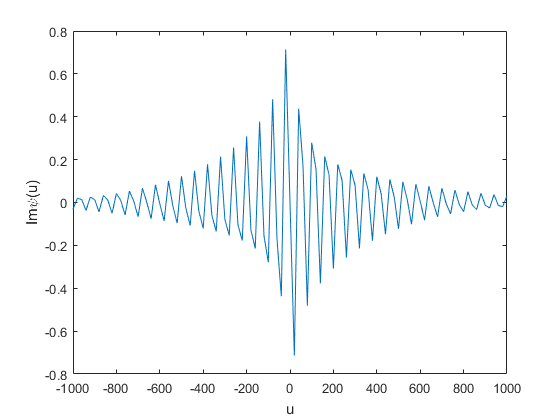}}}
 \caption{The real part of the function $\varphi_{Z_t}(u)$ under an IG subordinator (top) and its imaginary part(bottom). Parameters are obtained from estimating procedures for future oil prices as explained in  Section \ref{estimationAndOutput}. }\label{fig:subim2}
\end{center}
\end{figure}
 When the subordinator $L^j$ is a Gamma process with shape parameter $\alpha_j$ and rate parameter $\beta_j$, we have:
\begin{equation}\label{eq:tc_gamma}
    \Psi_{Z^j}(u)=-\alpha_j \log \Big(1+\frac{\mu_j u +\frac{1}{2}i \sigma^2_j u^2)+\beta_j^2}{\beta_j}\Big),\alpha_j>0,\beta_j>0, j=1,2
\end{equation}
To have discounted prices being a martingale, according to equation (\ref{martingale_condition}):
\begin{align}\nonumber
    \psi^j_{Gamma}(-\im)&=t^{-1}\log\Bigg[\Bigg(1+\frac{\im\mu_j u-\frac{(\sigma_j)^2u^2}{2}}{\beta_j}\Bigg)^{-\alpha_jt}\Bigg]\Bigg|_{u=-\im}=r \nonumber \\
    \nonumber
\end{align}
leading to:
\begin{equation*}\label{}
   \mu_j= -\beta_j(\exp(-\frac{r}{\alpha_j})-1)+\frac{(\sigma_j)^2}{2}
\end{equation*}
When the process $Z^j$ is a time-changed process subordinated by an Inverse Gaussian process with parameters $\alpha_j, \beta_j$, the characteristic function is given by equation (\ref{eq:tc_ig}) and therefore for each state $j\in E$ we solve for $\mu_j$:
\begin{align}
    \psi^j_{IG}(-\im)&=t^{-1}\log\big[\exp(-\alpha_j t\Big(\sqrt{2(-\im\mu_j u+\frac{(\sigma_j)^2u^2}{2})+(\beta_j)^2}-\beta_j\Big))\big]\big|_{u=-\im}=r \nonumber
    \end{align}
    It leads to:
    \begin{eqnarray*}\label{mu^j_IG}
     \mu^j&=&\frac{1}{2}[(\beta_j-\frac{r}{\alpha_j})^2+(\beta_j)^2]+\frac{\sigma^2_j}{2}.
    \end{eqnarray*}
   Holding all the other parameters constant, the drift verifies equation (\ref{martingale_condition}).
The value of $\mu_j$ is such that the $j$-th discounted price process, when the subordinator is an Inverse Gaussian process, is a martingale.
\end{example}

\section{Pricing European options under switching Levy models } \label{pricingMethods}
We turn back to pricing, consider a European call contract with maturity at $T$ and strike price $K$. Its payoff, written in terms of the log-returns, is:
\begin{equation*}\label{}
  h(Z_T)=(S_0 e^{Z_T}-K)_+=K (exp(x_0+Z_T)-1)_+
\end{equation*}
 where  $x_0:=\log(S_{0}/K)$.\\
  The price of the contract at a time $t<T$ and $x= \log (\frac{S_t}{K})$
  is denoted as $v(x,t)$ and verifies:
    \begin{equation}\label{eq:Fourier_COS_equation}
     v(x,t)=e^{-r\Delta t}~\mathbb{E}_{\theta}[v(y,T)]=e^{-r\Delta t}\int_\R v(y,T)f_{Z_T}(y|x)dy,
 \end{equation}\
 Notice that $v(y,T)$ is the payoff at maturity time $T$ and  $y:=\log(S_T/K)$.\\
  The value $\Delta t=T-t$ is the time to maturity and $r$ is the risk-neutral interest rate.  The function $f_{Z_T}(y|x)$ is the probability density function (p.d.f.) of $Z_T$ given the value $x=\log(S_{0}/K)$.\\
$\mathbb{E}_{\theta}$ is the expectation value operator with respect to an  EMM $\mathcal{Q}^{\theta}, \theta \in \mathbb{R}$ which is determined by an Esscher transform of the historic measure $P$. See  \cite{EsscherTransform} for a rationale in terms of a utility-maximization criteria. \\
For a stochastic process $(X_t)_{t \geq 0}$ the latter is defined as:
  \begin{equation}\label{eq:esscher}
  \frac{d \mathcal{Q}^{\theta}_t}{d P_t}=\exp(\theta X_t-t ~l_X(\theta)),\; 0 \leq t \leq T,\; \theta \in \mathbb{R}
\end{equation}
   where $P_t$ and $\mathcal{Q}^{\theta}_t$ are the respective restrictions of $P$ and $\mathcal{Q}^{\theta}$ to the $\sigma$-algebra $\mathcal{F}_t$. We define by $\varphi^{\theta}_{X_t}$, $\Psi^{\theta}_{X_t}$ and $ l^{\theta}_X(u)$  respectively the characteristic function, characteristic exponent and  moment generating function  of a process $(X_t)_{t \geq 0}$ under the probability $\mathcal{Q}^{\theta}$ obtained by an Esscher transformation as given in equation (\ref{eq:esscher}). \\
We follow a pricing approach via  Fourier- Cosine Series expansion of the p.d.f. $f_{Z_T}$. The  method has been proposed in \cite{Fourier}. \\
The solution to (\ref{eq:Fourier_COS_equation}) is obtained by expanding the p.d.f. $f_{Z_T}(./x)$  in terms of a Fourier basis under Gamma and Inverse Gaussian subordinators introduced in the previous section .\\
The domain of integration is truncated to a finite interval $[a,b]$ for the purposes of numerical integration, for a discussion about selecting the  truncation interval and its associated error, see \cite{convergence_COS}. \\
The Fourier-cosine expansion of $f_{Z_T}$ is given by:
 \begin{equation}\label{function_ab}
     f_{Z_T}(y|x)=\sum_{k=0}^{\infty}A_k(x)\cos \Big(k\pi\frac{y-a}{b-a}\Big),
 \end{equation}
 where the first term of the summation is weighted by one-half.\\
  The coefficients of the Fourier expansion, denoted by $A_k(x)$ , are approximated by:
\begin{equation*}
A_k(x)=\frac{2}{b-a}\text{Re}\Big\{\varphi_{Z_T} \Big(\frac{k\pi}{b-a}/x \Big) \exp\Big(-\im k\pi\frac{a}{b-a}\Big)\Big\}.
\end{equation*}
Hence, substituting (\ref{function_ab}) into equation (\ref{eq:Fourier_COS_equation}) we have:
  \begin{equation*}
     v(x,t)=\frac{1}{2}(b-a)e^{-r\Delta t}\sum_{k=0}^\infty A_k(x)V_k,
 \end{equation*}
 where $V_k$ is the Fourier coefficients of $v(y,T)$ given by:
 \begin{equation*}\label{eq:V_k}
 V_k:=\frac{2}{b-a}\int_a^b v(y,T)  \cos\Big(k\pi\frac{y-a}{b-a}\Big)dy.
 \end{equation*}
 In particular, for a European call option we have:
 \begin{equation*}
     V^{Call}_k=\frac{2 K}{b-a}\int_0^b (e^y-1)\cos\Big(k\pi\frac{y-a}{b-a}\Big)dy.
 \end{equation*}
 For a European put option denoted by $V^{Put}_k$ a similar expression is found.
 Finally, truncating the infinite series to $N$ terms, we obtain:
 \begin{equation*}\label{Characteristic}
 v(x,t)\approx e^{-r\Delta t} \sum_{0 \leq k < N} \text{Re} \Big\{\varphi_{Z_T} \Big(\frac{k\pi}{b-a}/x\Big) e^{-\im k\pi\frac{a}{b-a}} \Big\}V_k.
\end{equation*}

 The characteristic function of the log-return at maturity $Z_T$ is computed above by equation (\ref{eq:charfunc}) under  Inverse Gaussian and Gamma subordinators. As Fourier-cosine series of entire functions converges exponentially, so $N$ does not be too large to obtain good approximations. For European call options, it is found that the price is not accurate and extremely sensitive to the values of $b.$ On the other hand, it is also found that for large values of $b$,  $V^{Call}_k$ diverges   while $V^{Put}_k$ converges quickly and varies little with changes in the left-end of the truncation interval $a$. We  therefore rely on the put-call parity which allows for the computation of the European call option using the put option.

We summarize the calculations as follows:\\
\textbf{Algorithm}
\begin{enumerate}
    \item Initialization: choose appropriate boundary points $a,b$, number of terms $N$ in the series expansion and contract specifications (i.e.  interest rate $r$, initial stock price $S_0$, strike price $K$, with $x:=S_0/K$ and time to maturity $\Delta t$).
    \item Initialize $N\times 1$ array of payoffs $v^{Put}$ and $v^{Call}$.
    \item For $k=0$ to $N-1$
    \begin{enumerate}
        \item    Define the $k$-th element of $v^{Put}$ to be:\\ $v^{Put}(k)=e^{-rT}\text{Re}\Big\{\varphi_{Z_{\Delta t}}(k\pi/(b-a); x) \exp(-\im k\pi(a/(b-a))\Big\}V^{Put}_k$.\\
                \item $v^{Call}(k)=v^{Put}(k)+S_0-Ke^{-rT}$ \quad\quad(put-call parity)
    \end{enumerate}
    \item $v^{Call}_{final}=\frac{1}{2}v(1)+\sum_{k=1}^{N-1}v(k)$\quad\quad (Summation)
\end{enumerate}
The algorithm is implemented in a desktop computer using MATLAB. We compare the European call payoff and running time under Monte Carlo simulation and Fourier-Cosine pricing in Table \ref{monte carlo COS comparison} for different maturities and strike prices.

 \begin{table}[h!]
\centering
\caption{Comparison of European Call option Payoffs using Monte Carlo Simulation and Fourier-Cosine Pricing, as well as their computational times.}
 \begin{tabular}{|c|c|c|c|c|}
 \hline
($T,K)$ & Monte Carlo & Running Time (sec.) & Fourier-Cosine & Running time (sec.) \\ [0.5ex]
 \hline
 $(1,1)$ & 18.9554 & 9.31 &19.0401&0.0234 \\
 \hline
 $(2,1)$ & 19.9612 & 17.54 & 20.3456&0.1433\\
 \hline
 $(1,2)$ & 17.9942 & 10.01 &18.2164&0.1339\\
 \hline
 $(2,2)$& 19.0166& 11.40 & 18.5523&0.193 \\
 \hline

\end{tabular}
\label{monte carlo COS comparison}
\end{table}
In the parametric set considered (the next section discusses how to estimate the parameters), the Fourier-cosine method is on average 100 times faster than a standard Monte Carlo approach, while producing similar level of precision.\\
On the other hand, it is found that the difference between pricing European call options using Monte Carlo and Fourier-Cosine pricing remains constant for different strike prices. The error however grows linearly for increasing maturity times.\\
To compare with the price obtained via Monte Carlo we discuss the simulation procedure employed.\\
It is well-known that the  Markov chain $\{ s_t \}_{t \geq 0}$ spends  independent and  exponentially distributed random times between regime transitions. The $k-$th switching time is determined by:
\begin{equation*}
    \tau_k=\sum_{j=1}^k \Delta \tau_j
\end{equation*}
where $\Delta\tau_j$ are the times between regime changes. The parameters of the exponential random variables alternate between $\lambda_{12}$ and $\lambda_{21}$. \\
The differential equation is approximated numerically through finite differences using the Euler-Maruyama Method. Hence, if at time $t$ the process $Z$ is under regime $j\in E$, the increment of the process $Z$ during the time interval $[t, t+\Delta t)$ is given by:
\begin{equation*}
    \Delta Z^j_t=\mu^{j}\Delta L^{j}_t+\sigma_j\sqrt{\Delta L^j_t}N(0,1).
\end{equation*}
Notice that, for small $\Delta t$,  the process remains under regime $j$ with a probability close to one.\\

\begin{figure}[H]
\begin{center}
\subfigure[]{
\resizebox*{7cm}{!}{\includegraphics{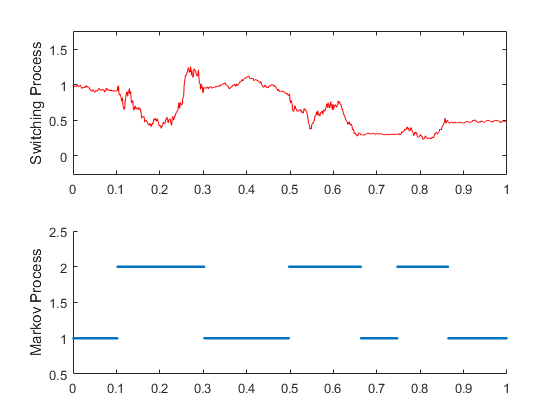}}}\hspace{5pt}
\subfigure[]{
\resizebox*{7cm}{!}{\includegraphics{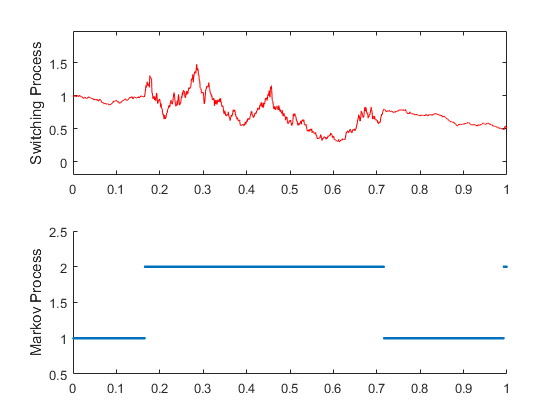}}}\hspace{5pt}
 \caption{Trajectories of a switching time-changed model. Parameters: $T=1, \mu_1=0.01, \mu_2=-0.1, \sigma_1=1, \sigma_2=5,  \alpha_1=\alpha_2=0.1, \beta_1=0.1, \beta_2=10, \lambda_{12}=5, \lambda_{21}=2. $, $\lambda_{12}=\lambda_{21}=4$, $\lambda_{12}=\frac{1}{10}, \lambda_{21}=4$}\label{traj 2}
\end{center}
\end{figure}
Figure  \ref{traj 2}(a) shows a single realization of a switching time-changed Levy process (top) under an IG subordinator, as well as the underlying  Markov chain (bottom). Figure  \ref{traj 2}(b) shows trajectories under a Gamma subordinator.\\
We devise an algorithm which computes the payoff of a European Call option by simulating $m$ independent realizations of the process $Z$ simultaneously and then estimating the price according to:
\begin{align*}
   v(x,T)  & \simeq \exp(-rT)\frac{1}{m}\sum_{j=1}^m (S_0 \exp(Z_{j,T})-K)_+
\end{align*}
where $Z_{j,T}$ is the $j$-th simulated log-price at maturity.\\
\begin{figure}[htb!]
\begin{center}
\subfigure[]{
\resizebox*{7cm}{!}{\includegraphics{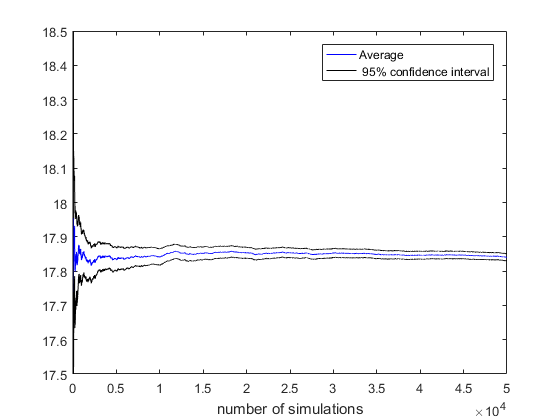}}}\hspace{5pt}
\subfigure[]{
\resizebox*{7cm}{!}{\includegraphics{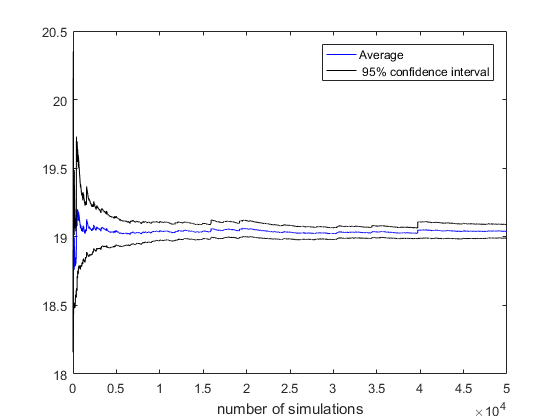}}}\hspace{5pt}
 \caption{Monte Carlo price and its confidence interval vs the number of simulations under  Gamma (top) and an IG (bottom) subordinators.  Parameters are the same than in previous figure except $\beta_1=0.1,\:\: \beta_2=0.01$.}\label{fig:confidence IG}
\end{center}
\end{figure}
We can also estimate the price using confidence intervals. The confidence interval is useful because it provides a range of values that are likely to contain the population mean. Endpoints of the confidence interval are given by:
\begin{equation*}
    \bar{h}\pm z_{0.95}\frac{s}{\sqrt{m}},
\end{equation*}
where $\bar{h}$ is the sample mean of the simulated payoffs, $s$ is the sample standard deviation, $m$ is the sample size and $z_{0.95}$ is the $95\%$ -percentile  of the normal distribution.\\
The confidence interval decreases as the number of simulations approaches infinity, however in the case of the Inverse Gaussian subordinator, the interval is larger at each simulation because Inverse Gaussian random variables have a larger variance than Gamma random variables  when $\beta<1$.\\
  At $m=10^4,$ the confidence interval when the subordinator is Inverse Gaussian is $ [18.7353,  19.1168]$. When the subordinator is a Gamma process, the confidence interval is $[ 17.8768,   17.9136]$.\\

\begin{figure}[htb!]
\begin{center}
\subfigure[]{
\resizebox*{7cm}{!}{\includegraphics{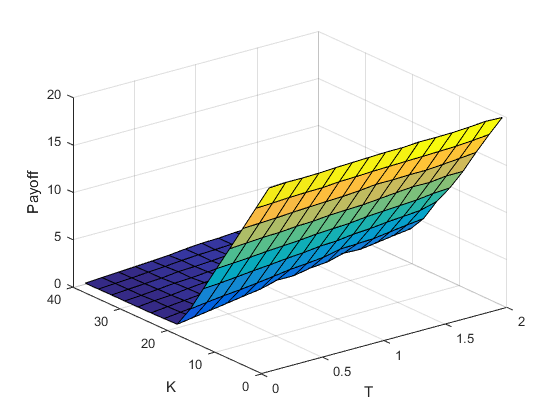}}}\hspace{5pt}
\subfigure[]{
\resizebox*{7cm}{!}{\includegraphics{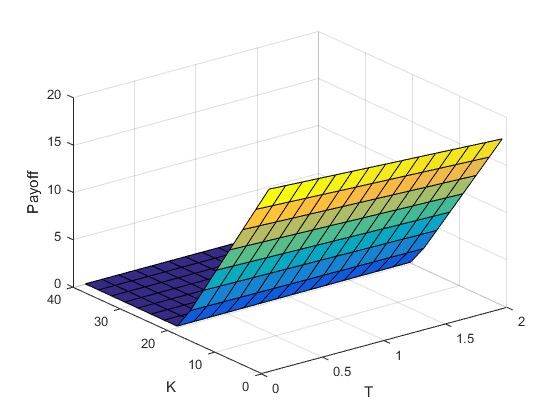}}}\hspace{5pt}
 \caption{ European call payoff as a function of $T$ and $K$ under two different subordinators with risk neutral drift. The other parameters are identical for each figure:  $\sigma_1=0.03, \sigma_2=0.7, \alpha_1=\alpha_2=0.1, \beta_1=1, \beta_2=1.2, \lambda_{12}=2.5, \lambda_{21}=1$. For an IG subordinator $\mu_1=0.3204, \mu_2=0.6450$. For a Gamma subordinator  $\mu_1=-0.2316, \mu_2=0.0541$.}\label{fig:image2}
\end{center}
\end{figure}
Figure  \ref{fig:image2} indicates the behaviour of the price of a European call option for a regime switching time-changed Levy process model under Inverse Gaussian (top) and Gamma subordinators (bottom). Both payoff models are monotone in $T$ and $K$. Moreover as $T$ increases, the expected payoff increases.  For $K>>S_0$ the probability that $S_T\geq K$ is very small, therefore the payoff is  close to $0.$\\
\begin{figure}[htb!]
\begin{center}
\subfigure[]{
\resizebox*{7cm}{!}{\includegraphics{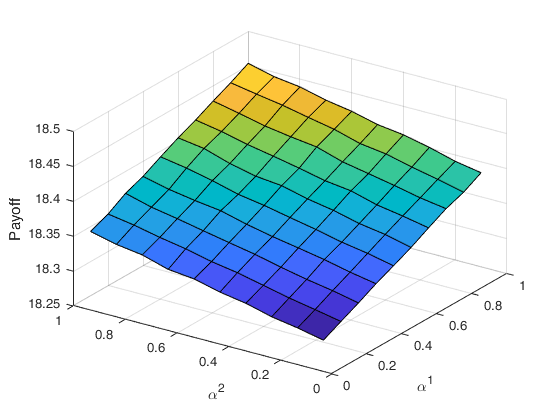}}}\hspace{5pt}
\subfigure[]{
\resizebox*{7cm}{!}{\includegraphics{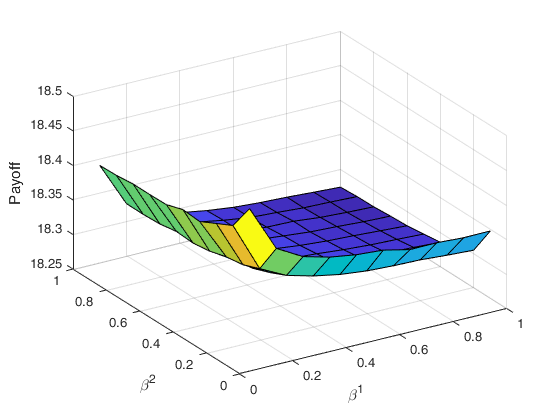}}}\hspace{5pt}
 \caption{Payoff as a function of parameters $\alpha_1, \alpha_2$. Payoff as a function of parameters $\beta_1, \beta_2$}\label{fig:G_b1_b2}
\end{center}
\end{figure}

\begin{figure}[htb!]
\begin{center}
\subfigure[]{
\resizebox*{7cm}{!}{\includegraphics{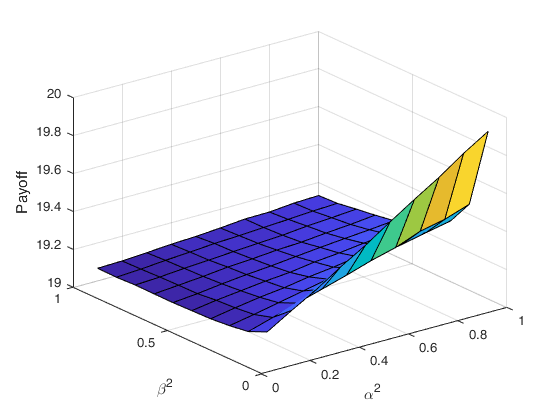}}}\hspace{5pt}
\subfigure[]{
\resizebox*{7cm}{!}{\includegraphics{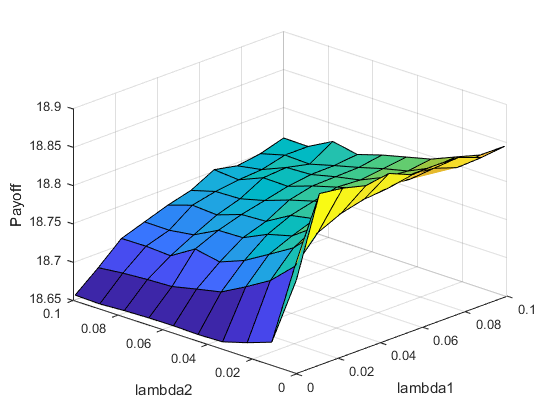}}}\hspace{5pt}
 \caption{ Payoff as a function of parameters $\alpha_2, \beta_2$. Payoff as a function of parameters $\lambda_{12}, \lambda_{21}$}\label{fig:image3}
\end{center}
\end{figure}

 In each figure parameters between states are held constant, except the ones indicated in the graph: $\mu_1=\mu_2=0.01,\:\: \alpha_1=\alpha_2=\beta_1=\beta_2=0.1,\:\: \sigma_1=\sigma_2=0.01,\:\lambda_{12}=\lambda_{21}=0.25,\: r=0.04,\: T=1,\: S_0=20\:$ and $ K=1.$ Setting the parameters $\lambda_{12}=\lambda_{21}$ implies the process spends an equal amount of time in each regime, on average. The only exception made is in Figure \ref{fig:G_b1_b2}, where $\lambda_{12}=1000$ and $\lambda_{21}=1/10$ so that the process would spend a majority of the time in the second regime; this makes the process an approximation to a time-changed Levy process.\\
  For changes in $\beta_1, \beta_2,$ the payoff approaches an asymptote because Gamma and Inverse Gaussian random variables both have mean $\alpha/\beta,$ which approaches infinity for $\beta\rightarrow 0^+.$
In Figure \ref{fig:image3} changes in the price  with respect to the intensity parameters and the parameters of the underlynig subordinators are shown.

 \begin{table}[h!]
\centering
\caption{European call option payoff comparison between the Black Scholes formula and Monte Carlo simulation of the reduced switching Levy process at different parameters}\label{bshpri}
\begin{tabular}[H]{|c|c|c|}
 \hline
 Parameters &  Reduced Switching Levy Model&Black Scholes formula
\\
$(T,K,r,\sigma)$&   $(\#\:\: \text{simulations}\:\: N=10^6)$&
	\\
 \hline
 (1,1,0.04,0.5)	& 19.0463	&19.0392\\
 \hline
(3,1,0.1,1)&	19.2955	& 19.3139 \\
\hline
 (2,30,0.5,0.001)	&8.963608	& 8.96361\\
\hline
 \end{tabular}
 \end{table}

 Notice that we can reduce the regime switching Levy processes to the Black-Scholes model  by defining the subordinator $L_t=t$ and setting the parameters equal across both regimes. Setting the drift  such that the discounted Black Scholes price process is a martingale $\mu_1=r-\frac{1}{2}\sigma^2_1$.
 we find that the price by a Fourier-cosine approach  is consistent with the price given by the  Black-Scholes formula. See Table \ref{bshpri}.

 \begin{figure}[htb!]
\begin{center}
\subfigure[]{
\resizebox*{7cm}{!}{\includegraphics{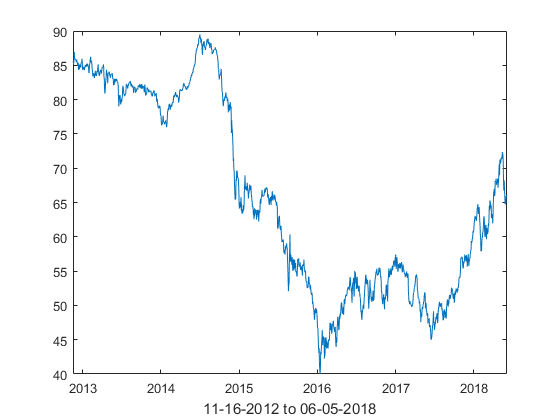}}}
\subfigure[]{
\resizebox*{7cm}{!}{\includegraphics{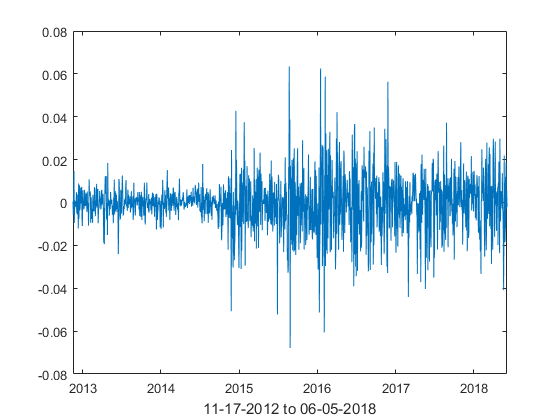}}}\hspace{5pt}
 \caption{Daily price series  and log-returns for WTI futures (left) and log-returns of WTI futures in NYMEX.
Source: Blooomberg Terminal, April 2018 }\label{fig:Estimation1}
\end{center}
\end{figure}

 \begin{figure}[htb!]
\begin{center}
\subfigure[]{
\resizebox*{7cm}{!}{\includegraphics{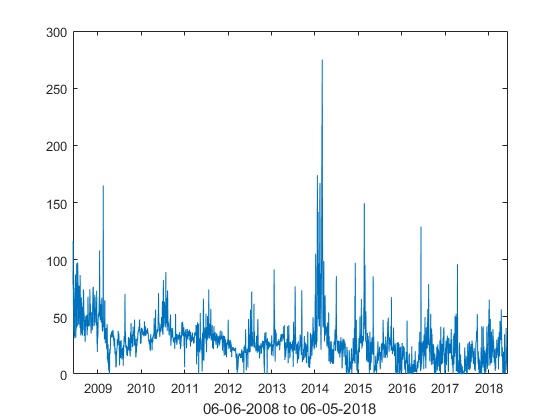}}}
\subfigure[]{
\resizebox*{7cm}{!}{\includegraphics{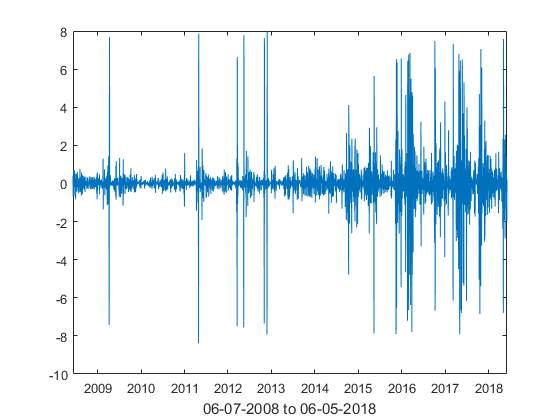}}}\hspace{5pt}

 \caption{Price series and log-returns for Ontario daily average electricity spot price.
Source: Blooomberg Terminal, April 2018.}\label{fig:Estimation2}
\end{center}
\end{figure}

\section{Parameter estimation and numerical pricing} \label{estimationAndOutput}
\begin{figure}[htb!]
\begin{center}
\subfigure[]{
\resizebox*{7cm}{!}{\includegraphics{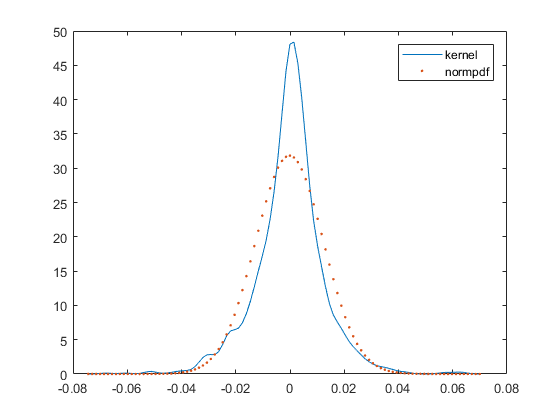}}}
\subfigure[]{
\resizebox*{7cm}{!}{\includegraphics{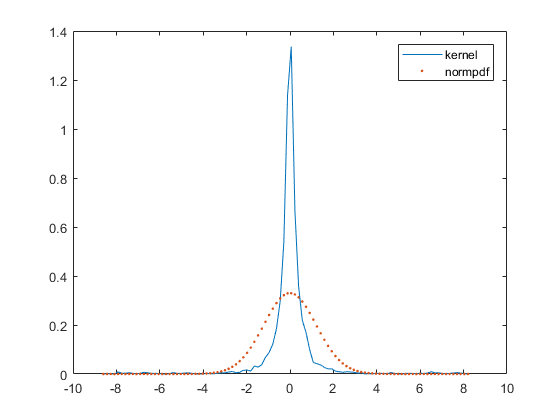}}}\hspace{5pt}
 \caption{Empirical density functions vs. normal distributions
oil log-returns.electricity log-returns}\label{fig:image5}
\end{center}
\end{figure}

We implement two approaches of fitting the parameters of the underlying model to financial historical data: calibration and estimation depending when  option's prices or the underlying electricity and oil prices are considered. In a calibration approach, the parameters are fitting by minimizing the quadratic error between the prices obtained numerically and option quotes. The option quotes are taken from Bloomberg's data base for a variety of strike prices and times to maturity.  The parameters are fitted using  European call option quotes of West Texas Intermediate (WTI) crude oil. \\
 In a parameter estimation, we implement the following techniques based on historical prices of oil and electricity: method of moments, minimum distance method and maximum likelihood estimation, combined with empirical estimation of the switching parameters. Specifically, we use daily historical NYMEX WTI crude oil futures (11-16-2012 to 06-05-2018) and IESO Ontario (Canada) Zone 24H electricity average spot prices (06-06-2008 to 06-05-2018). Spikes and stochastic volatility are observed in the series, Similar phenomena have been reported in other electricity markets, see \cite{benth1, cartea}. We assume that there are 250 trading days in a year with each trading day corresponding to $\Delta t=1/250$ of unit time. \\
Figures \ref{fig:Estimation1} and \ref{fig:Estimation2} plot the historic WTI oil futures and average Ontario electricity prices, as well as their log-returns. Electricity spot prices sometimes move below zero, implying a surplus of electricity produced during low demand. Because  electricity produced by power suppliers must be consumed immediately, the supplier pays wholesale customers to buy the surplus energy, see \cite{Electricity_Regime}. All negative prices are arbitrarily modified to CAD $\$ 0.01$ for estimation purposes.\\
We also compare the empirical density function of the log-returns of each commodity to the normal distribution under the same mean and variance parameters as the historical log-return process. To this end we implement a kernel smooth technique. Hence, the estimated p.d.f. is:
 \begin{equation*}\label{eq:kerel}
     \hat{f}(x; \theta)=\frac{1}{nh}\sum_{j=1}^n K\Big(\frac{x-z_j}{h}\Big),
 \end{equation*}
 where the function $K$ is the Gaussian kernel.\\
As the p.d.f. $f(x_k;\theta)$ of the log-returns is not available in a close form we simulate the model under a parametric set $\theta$  to get the empirical p.d.f. $\hat{f}(x_k;\theta)$ using  a kernel smoothing technique, see for example \cite{kernel}, and continue the optimization procedure. \\
The value of $h,$ is chosen to equal Silverman's quantity $h=1.06 \sigma n^{-1/5}$, where $\sigma$ is the standard deviation of the log-return series, see \cite{MaxLikelihood}. See Figure \ref{fig:image5} for empirical p.d.f.'s corresponding to WTI futures (left) and Ontario electricity log-return prices(right).\\
  The kurtosis of WTI future and Ontario electricity log-return series are respectively 6.34 and 23.947,  much larger than that of the normal distribution, suggesting the presence of heavier tails and extreme behavior. Significant negative skewness is also reported on both series, respectively -0.1314 and -0.1377\\
 In our model the parameters are described by  vectors $ \theta^j=(\mu_j,\sigma_j,\alpha_j,\beta_j)$ for $j=1,2,$ while $\lambda_{12}$ and $\lambda_{21}$ reflect the parameters of the hidden Markov chain. Hence the times the chain  remains in regime $j$ are independent and exponentially distributed with mean $\lambda^j=\frac{1}{\lambda_{jk}}$, with $k=\mod(2)+1$. We set $\Theta^j\subset\R^4$ to be the set of all feasible parameters for $\theta^j$. We assume that the two sets of parameters belong in different parameter spaces i.e. $\Theta^1\neq\Theta^2.$ The two parameters of the subordinator and the diffusion coefficient are required to be positive, therefore we add the natural constraints $\sigma_j>0, \alpha_j>0, \beta_j>0.$\\
 We assume that the duration of the $j$-th observed historic regime is the most probable value i.e. it is equal to the expectation value $\lambda^j.$ For each commodity, the $j$-th holding-rate parameter is given by:
 \begin{equation*}
     \lambda^j=\frac{\text{total number of days in regime}\:\: j }{\text{ number of occurrences of regime}\:\: j}
 \end{equation*}
 where it is assumed that there are 250 trading days every year.\\
 By simple inspection of the log-return process data of oil futures we set the process to be in regime one between 11-16-2012 and  11-16-2014 as well as between 02-06-2017 and  06-05-2018; otherwise, we assume that the process is in regime two.  In the case of the log-returns of electricity spot prices; we set the process be in regime two whenever the absolute value of the log-returns exceeds 3 and in regime one otherwise.  \\
Table \ref{tab:lambda} shows the average time and daily standard deviation in the two regimes of the WTI futures and Ontario electricity series.   We also include the variance of the log-returns within each regime; the different orders of magnitude between regimes justifies the use of a switching model.\\
 \begin{table}[h]
     \centering
     \begin{tabular}{|c|c|c|c|c|c}
       \hline
          Commodity& $\hat{\lambda}^1$&$\hat{\lambda}^2$ & St. Dev. (regime 1)& St. Dev. (regime 2)\\
          \hline
          Oil&0.900&3.80&0.0052&0.0148\\
          \hline
          Electricity & 0.2618&0.0081&0.6020&6.1086 \\
          \hline
     \end{tabular}
     \caption{Holding-rate parameters estimation for each commodity as well as the variance in each regime}
     \label{tab:lambda}
 \end{table}
 By having defined the location of the regime changes and therefore estimated the values of $\lambda^1, \lambda^2$, the historic log-returns are separated into two sets of data, one containing all the data points for each regime.\\
  To calibrate the parameters  within each regime we minimize the \textit{mean square error} between the numeric option payoffs and European call option quotes. When the option is out of the money the option price is obtained using a Monte Carlo approach because the Fourier Cosine method exhibits significant error.\\
  Thus, the objective function in regime $j$ is
 \begin{align*}
     J(\theta^j)&=\sqrt{\frac{1}{n}\sum_{T,K}(V(T,K; \theta^j)-\hat{V}(T,K; \theta^j))^2},\quad j\in\{1,2\}.
 \end{align*}
 where, by a convenient change in notation to emphasize the dependence on the parameters we write  $V(T,K; \theta^j)$ and the option quotes $\hat{V}(T,K; \theta^j)$,  taken  over a range of strike prices $K$ and maturity times $T$.\\
The optimal parameter is $\hat{\theta}^j=\operatorname*{arg\,min}_{\theta\in\Theta^j} J(\theta^j)$. It is calculated using a  gradient descent method.\\
The stopping criteria is taken to be step tolerance, taken to be equal to 1e-10. The k-th step tolerance is a lower bound on the size of the step $||\theta^t-\theta^{t-1}||_2$. The solver stops if the stopping criteria is reached, or if the maximum number of iterations (fixed to 1000 steps) is exceeded.  Different initial starting points are found to give similar estimation of the parameters. Table \ref{tab:calibration} gives the estimated calibration in the case when the subordinator is a Gamma process or an Inverse Gaussian process. As expected, the volatility is higher in regime two in the case of both subordinators.

\begin{table}[h!]
\centering
\caption{Parameter Calibration using a Mean Square Error criteria}
\begin{tabular}{|c|c|c|c|c|}
\hline

Commodity (subordinator) & $\hat{\mu}$ & $\hat{\sigma}$ & $\hat{\alpha}$ & $\hat{\beta}$\\
 \hline
Oil log-return Regime 1 (Gamma)&  -0.03387 &  0.0030&   2.640710  & 1.007e-8\\
 \hline

Oil log-return Regime 2 (Gamma)& -0.01445&   1.116184&  2.56567e-5 &  10.32441 \\
 \hline
Oil log-return 1 (Inverse Gaussian) & -0.04976&   0.130011&   0.24788 & 92.6926\\
\hline

Oil log-return 2 (Inverse Gaussian) &   -0.04950&  0.515891&   8.531e-4&  8.43091 \\ \hline
 \end{tabular}
 \label{tab:calibration}
 \end{table}
To choose the initial set of parameters we use the \textit{Method of Moments}. Theoretical moments are computed from the characteristic function of the model under both subordinators considered. Matching both, empirical and theoretical moments up to order forth leads to the following non-linear system of equations, in the case of a model under an Inverse Gaussian subordinator:
 \begin{align*}
{\hat\mu}_1&={\alpha} {\mu} {\Delta t}/{\beta}\\
\hat{\mu}_2 &={\alpha} {\Delta t} ({\mu}^{2} + {\beta}^{2} {\sigma}^{2} + {\alpha} {\beta} {\mu}^{2} {\Delta t})/{\beta}^{3}\\
\hat{\mu}_3&={\alpha} {\mu} {\Delta t} ({3} {\mu}^{2} + {3} {\beta}^{2} {\sigma}^{2} + {3} {\alpha} {\beta} {\mu}^{2} {\Delta t} + {\alpha}^{2} {\beta}^{2} {\mu}^{2} {\Delta t}^{2} + {3} {\alpha} {\beta}^{3} {\sigma}^{2} {\Delta t})/{\beta}^{5}\\
\hat{\mu}_4&=({\alpha} {\Delta t} ({15} {\mu}^{4} + {3} {\beta}^{4} {\sigma}^{4} + {18} {\beta}^{2} {\mu}^{2} {\sigma}^{2} + {15} {\alpha} {\beta} {\mu}^{4} {\Delta t} + \\ &\quad\quad{6} {\alpha}^{2} {\beta}^{2} {\mu}^{4} {\Delta t}^{2} + {\alpha}^{3} {\beta}^{3} {\mu}^{4} {\Delta t}^{3} + {3} {\alpha} {\beta}^{5} {\sigma}^{4} {\Delta t} + {6} {\alpha}^{2} {\beta}^{4} {\mu}^{2} {\sigma}^{2} {\Delta t}^{2} + {18} {\alpha} {\beta}^{3} {\mu}^{2} {\sigma}^{2} {\Delta t}))/{\beta}^{7}.
\end{align*}
where ${\hat\mu}_k$ is the empirical k-th moment.\\
The system of equations is solved separately for each regime  using the function \verb fsolve in MATLAB based on the  trust region algorithm.  The results are summarized in Table \ref{tab:method_moment IG}.\\
A similar result is obtained for the model under a Gamma subordinator.\\
 \begin{table}[h!]
\centering
\caption{Parameter Estimation using Method of Moments under Inverse Gaussian Subordinator}
\begin{tabular}{|c|c|c|c|c|}
\hline

Commodity & $\hat{\mu}$ & $\hat{\sigma}$ & $\hat{\alpha}$ & $\hat{\beta}$\\
 \hline
Oil log-return Regime 1& 0.1624  &  0.7213  &  0.3238  &  1.6971\\
 \hline

Oil log-return Regime 2 & -0.0354  &  1.3402  &  0.0400  &  1.9584\\
 \hline
Electricity log-return 1  &	 0.1111  &  3.1233 &  28.4386  &  3.4862
\\
\hline

Electricity log-return 2 &  -3.7405 &  19.5346  &  0.0132  &  0.3539\\
\hline

 \end{tabular}
 \label{tab:method_moment IG}
 \end{table}
The method encounter difficulties to find a global minimum in the case where the empirical moments were calculated using electricity log-return prices. Changing the initial starting points resulted in varying results, which indicates the presence of local minima. Despite these shortcomings the method of moments can be used as an initial solution for a \textit{Minimum Distance} approach based on the distance between the theoretical and empirical characteristics function of the log-returns, the later defined as:
\begin{equation*}\label{eq:CHAR}
    \hat{\varphi}_{Z^j_{\Delta t}}(u)=\frac{1}{n}\sum_{k=1}^n\exp(\im u z_k)
\end{equation*}
for a sample $z_1, z_2, \ldots, z_n$ of $n$ log-returns of the underlying series. See \cite{MinimumDistanceEstimates2} for details.\\
The objective function under regime $j$ is defined by:
\begin{equation*}
    ||\varphi_{Z^j_{\Delta t}}(u; \theta)- \hat{\varphi}_{Z^j_{\Delta t}}(u)||_2 :=\Bigg(\int_{-\infty}^{\infty} |\varphi_{Z^j_{\Delta t}}(u; \theta)- \hat{\varphi}_{Z^j_{\Delta t}}(u)|^2 w(x)dx\Bigg)^{1/2},
\end{equation*}
where $w$ is the weight function  $w(x)=(1/\sqrt{2\pi})\exp(-x^2/2)$.\\
Then, $\hat{\theta}$ is the minimum distance estimate of $\theta$ if
\begin{equation*}
 ||\varphi_{Z^j_{\Delta t}}(u; \hat{\theta})- \hat{\varphi}_{Z^j_{\Delta t}}(u)||_2  =\inf_{\theta\in\Theta}\{ ||\varphi_{Z^j_{\Delta t}}(u; \theta)- \hat{\varphi}_{Z^j_{\Delta t}}(u)||_2 \}
\end{equation*}
Again, we apply the algorithm to each regime separately.\\
The integral is computed numerically using a global adaptive quadrature algorithm, where the interval of integration is subdivided and the integration takes place on each subdivided interval. Intervals are further subdivided if the algorithm determines that the integral is not computed to sufficient accuracy.
 \begin{table}[h!]
\centering
\caption{Parameter Estimation using Minimum Distance Method under Inverse Gaussian subordinator}
\begin{tabular}{|c|c|c|c|c|}
\hline

Commodity & $\hat{\mu}$ & $\hat{\sigma}$ & $\hat{\alpha}$ & $\hat{\beta}$\\
 \hline
Oil log-return Regime 1& 0.01736&   0.11675& 31.648&   8.0554\\
 \hline

Oil log-return Regime 2 & -0.4956&   2.0078&   2.2260&  10.141\\
 \hline
Electricity log-return 1  &0.00813&  02.0139&   67.456 &  0.00154
\\
\hline

Electricity log-return 2 & 5.7435 & 4.48714& 76.004 &  6.871e-4\\
\hline

 \end{tabular}
 \label{tab:Min dist method_IG}
 \end{table}
In table \ref{tab:Min dist method_IG} estimates of the model parameters under both regimes and for the two series of underlying assets are shown.\\
Given a random sample $x=(x_1, ..., x_n)$ of a random variable $X$ with an associated density function $f(x;\theta)$ of the data $x$ under the real world and unknown parameters  $\theta$, maximum likelihood estimation (MLE) is a method used to estimate the vector valued parameter $\theta$ of the model by maximizing the likelihood function:
\begin{equation*}
    l(\theta;z)=\sum_{k=1}^n \log f(z_k;\theta);\:\:\:\: \theta\in\Theta,
\end{equation*}
with respect to $\theta$. The value of $\theta$ is constrained to $\Theta\subset\R^4$, the space of all feasible values of the parameters. The maximum likelihood function $\mathcal{L}$ is primarily a function of the unknown parameters $\theta$.  The maximum likelihood estimator is given by:
\begin{equation*}\label{maxlikelihoodeq}
    \hat{\theta}=\arg\max_{\theta\in\Theta}l(\theta;x)
\end{equation*}
In tables \ref{mlegamma} and \ref{mleig} estimations  based on the empirical m.l.e. for both regimes and models under Gamma and Inverse Gaussian subordinators are shown.
\begin{table}[h!]
\centering
\caption{Parameter Estimation using Maximum Likelihood Method under Gamma subordinator}\label{mlegamma}
\begin{tabular}{|c|c|c|c|c|}
\hline

Commodity & $\hat{\mu}$ & $\hat{\sigma}$ & $\hat{\alpha}$ & $\hat{\beta}$\\
 \hline
Oil log-return Regime 1& 0.0023& 0.0431&  42.928&  11.9960\\
 \hline

Oil log-return Regime 2 & -0.372&   0.52851 & 17.3008& 88.556\\
 \hline
Electricity log-return 1  & 5.844e-3 &  1.5002 & 93.271  & 2.1903\\
\hline

Electricity log-return 2 & -0.0148&  7.543&  90.5900&   0.01770
 \\
\hline

 \end{tabular}
 \end{table}
\begin{table}[h!]
\centering
\caption{Parameter Estimation using Maximum Likelihood Method under Inverse Gaussian subordinator}\label{mleig}
\begin{tabular}{|c|c|c|c|c|}
\hline
Commodity & $\hat{\mu}$ & $\hat{\sigma}$ & $\hat{\alpha}$ & $\hat{\beta}$\\
 \hline
Oil log-return Regime 1&-0.4883&   0.5058&   0.64603&  63.709 \\
 \hline
 Oil log-return Regime 2 & 0.1201  & 2.9707  & 0.00014 &  9.993\\
 \hline
Electricity log-return 1  &-0.1781&  0.25873&  0.91878& 20.0860\\
\hline

Electricity log-return 2 &  -0.0191&   4.9752 &  5.512e-5&  11.016\\
\hline
 \end{tabular}
 \end{table}
 In each  case, the values of the volatility  $\sigma$ are found to be higher in the second regime, hence justifying the use of a regime switching model. In nearly every method, the value of $|\mu|$ was found to be very small, which is expected as the long term deterministic contribution to the process is expected to be near zero.\\
 In choosing constraints, we set the lower bound of $\sigma, \alpha, \beta $ to be some small number $\epsilon=10^{-6}$. We set the upper bound of $\sigma$ to 5 as the diffusion is  expected to be smaller than 1 and for $\alpha,\:\: \beta,$ we set the upper bound to be 100, as the expected value of both Inverse Gaussian and Gamma random variables depends on the ratio $\alpha/\beta$ rather than any particular value for $\alpha$ and $\beta$. The drift $\mu$ is expected to be small, so in most cases, it was constrained to the set $[-1,1]$.
\section{Conclusions}
We price European-style options with oil and electricity prices as underlying assets under a switching Levy time-changed noise. These are realistic models that allow to incorporate stylized features in the dynamic of the prices. Our findings show that under this framework a pricing method based on a Fourier-cosine offers an efficient and accurate result when compared with a standard Monte Carlo approach.\\
In addition, we address the problem of parameter estimation and calibration. To this end we successfully tried different methods based on both, historic and risk-neutral measure.
\section{Acknowledgments}
This research is supported by the Natural Sciences and Engineering Research Council of Canada.

\end{document}